\documentclass[useAMS,usenatbib]{mn2e}
\usepackage{color}
\usepackage{tabularx}
\usepackage{multirow}
\usepackage{amsmath}
\usepackage[para,online,flushleft]{threeparttable}
\usepackage[dvipdfmx]{graphicx}

\usepackage{color}

\title[Off-axis jet scenario for GRB 190829A]{
Off-axis jet scenario for early afterglow emission of low-luminosity gamma-ray burst GRB 190829A
}
\author[Sato et al.]
{Yuri~Sato$^{1}$\thanks{E-mail: yuris@phys.aoyama.ac.jp (YS)},
Kaori~Obayashi$^{1}$\thanks{E-mail: o-kaori@phys.aoyama.ac.jp (KO)}, 
Ryo~Yamazaki$^{1,2}$\thanks{E-mail: ryo@phys.aoyama.ac.jp (RY)},
Kohta~Murase$^{3,4,5,6}$\thanks{E-mail: murase@psu.edu (KM)}
\and
and Yutaka~Ohira$^{7}$\thanks{E-mail: y.ohira@eps.s.u-tokyo.ac.jp (YO)}\\
$^{1}$Department of Physics and Mathematics, Aoyama Gakuin University, 5-10-1 Fuchinobe, Sagamihara 252-5258, Japan\\
$^{2}$Institute of Laser Engineering, Osaka University, 2-6 Yamadaoka, Suita, Osaka 565-0871, Japan\\
$^{3}$Department of Physics, Pennsylvania State University, University Park, Pennsylvania 16802, USA\\
$^{4}$Department of Astronomy \& Astrophysics, Pennsylvania State University, University Park, Pennsylvania 16802, USA \\
$^{5}$Center for Multimessenger Astrophysics, Pennsylvania State University, University Park, Pennsylvania 16802, USA\\
$^{6}$Center for Gravitational Physics, Yukawa Institute for Theoretical Physics, Kyoto University, Kyoto, Kyoto 606-8502, Japan\\
$^{7}$Department of Earth and Planetary Science, The University of Tokyo, 7-3-1 Hongo, Bunkyo-ku, Tokyo 113-0033, Japan
}
\begin{document} 
\date{today}
\pagerange{\pageref{firstpage}--\pageref{lastpage}} \pubyear{2021}
\maketitle
\label{firstpage}

\begin{abstract}
Recently, ground-based Imaging Atmospheric Cherenkov Telescopes have reported the detection of very-high-energy (VHE) gamma-rays from some gamma-ray bursts (GRBs).
One of them, GRB~190829A, was triggered by the {\it Swift} satellite, and about $2\times10^4$~s after the burst onset
the VHE gamma-ray emission was detected by H.E.S.S. with $\sim5\sigma$ significance.
This event had unusual features of having much smaller isotropic equivalent gamma-ray energy than typical long GRBs and achromatic peaks in X-ray and optical afterglow at about $1.4\times10^3$~s.
Here we propose an off-axis jet scenario that explains these observational results. 
In this model, the relativistic beaming effect is responsible for the apparently small isotropic gamma-ray energy and spectral peak energy.
Using a jetted afterglow model, we find
that the narrow jet, which has the initial Lorentz factor of 350 and the initial jet opening half-angle of 0.015~rad, 
viewed off-axis can describe the observed achromatic behavior in the X-ray and optical afterglow.
Another wide, baryon-loaded jet is necessary for the later-epoch X-ray and radio emissions.
According to our model, the VHE gamma rays observed by H.E.S.S. at $2\times10^4$~s may come from the narrow jet
through the synchrotron self-Compton process.
\end{abstract}
 


\section{Introduction}

Recently, very-high-energy (VHE) gamma-rays from some gamma-ray bursts (GRBs) were detected by ground-based Imaging Atmospheric Cherenkov Telescopes, such as the Major Atmospheric Gamma Imaging Cherenkov (MAGIC) telescopes, 
and the High Energy Stereoscopic System (H.E.S.S.).
A prototypical example so far is GRB~190114C, which was simultaneously detected with MAGIC and {\it Fermi} Large Area Telescope \citep{Acciari2019,MAGIC2019,Ajello2020}.
The observed spectrum in the VHE gamma-ray band is well explained by the synchrotron self-Compton (SSC) model
\citep{MAGIC2019,derishev2019,fraija2019a,fraija2019b,fraija2019c,wang2019,Asano2020,HuangXL2020b}.
H.E.S.S. detected VHE gamma-rays from GRB~180720B about 10~hours after the burst onset at 5.3~$\sigma$ significance level, and the  energy flux was $\nu F_\nu\approx5\times10^{-11}$erg~s$^{-1}$cm$^{-2}$ in
the VHE band \citep{abdalla2019}.
GRB~190829A was also detected by H.E.S.S. about $2\times10^4$~s after the burst trigger \citep{Naurious2019}.
Its significance is $\sim5~\sigma$. 
Moreover, a possible detection of VHE gamma-ray emission from a short GRB~160821B has been claimed by MAGIC \citep{MAGIC2021,Zhang2021a}.
It is expected that  in the near future, the Cherenkov Telescope Array \citep[CTA;][]{Actis2011} will increase the number of GRBs with VHE gamma-rays~\citep{kakuwa2012,gilmore2013,inoue2013}.

Compared with GRB~190114C and 180720B, GRB~190829A has some peculiar observational properties. The prompt gamma-ray emission (from $\sim10$~keV to MeV band) consists of two temporally separated components \citep{Chand2020}.
The burst started with less energetic emission \citep[hereafter Episode~1 following][]{Chand2020} with an isotropic equivalent gamma-ray energy of $E_{\rm iso,\gamma}=3.2\times10^{49}$~erg and a peak energy (that is, the photon energy at which the $\nu F_{\nu}$-spectrum takes a maximum) $E_p=120$~keV.
After quiescent time interval lasting about 40~s, the second brighter emission (Episode~2) with $E_{\rm iso,\gamma}=1.9\times10^{50} $~erg and $E_p=11$~keV, appeared.
The observed values of $E_{\rm iso,\gamma}$ and $E_p$ of Episode~2 are consistent with Amati relation
\citep{amati2002,Sakamoto2008}, 
while those of Episode~1 are in the region of low-luminosity GRBs.
Both Episode~1 and 2 have smaller $E_{\rm iso,\gamma}$ and $E_p$ than typical long GRBs, including the other VHE gamma-ray events, GRB~190114C and 180720B \citep[e.g.,][]{HuangXL2020}.
Indeed, GRB~190829A occurred so nearby with a redshift of  $z=0.0785$ that such weak prompt emissions could be observed.

Well-sampled afterglow light curves of GRB~190829A were obtained in X-ray, optical/infrared (IR) \citep{Chand2020},  and radio bands \citep{Rhodes2020}.
It is remarkable that early X-ray and optical/IR afterglow emission showed a rising part and simultaneously peaked at about $1.4\times10^3$~s.
Such an ``achromatic'' behavior is difficult to be explained in standard afterglow model, in which the synchrotron emission has the maximum when the typical frequency $\nu_m$ crosses the observation bands \citep{sari1998}.
In contrast, the other VHE events, GRB~190114C and 180720B, showed monotonically decaying X-ray afterglow emission
\citep{yamazaki2020,fraija2019b}.
Possible interpretations of the achromatic bump are the X-ray flare with optical counterpart
\citep{Chand2020,Zhang2020b,Zhao2020b} and the afterglow onset of baryon loaded outflow with bulk Lorentz factor of about 30 \citep{fraija2020a}.
Another interesting point is that late time ($t\ga10^{4-5}$~s) optical/IR emissions are dominated by supernova component \citep{Hu2020}.

\begin{figure*} 
\centering \vspace*{10pt}
\includegraphics[width=1.0\textwidth]{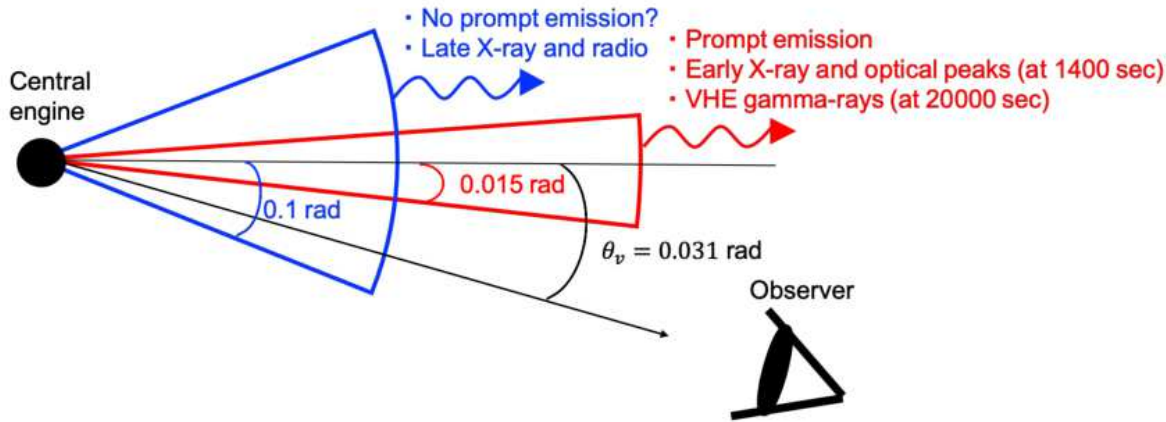}
\caption{
Schematic view of our two-component jet model for GRB~190829A. The red and blue cones represent narrow and wide jets, respectively.
Initial shapes of the jets are depicted with their initial opening half-angles.
The black arrow shows the observer's line of sight.
As the jets expand, they spread sideways, and at $\sim2\times10^4$~s when H.E.S.S. detected VHE gamma-rays, the observer's line of sight is inside the cone of the narrow jet. 
}
\label{fig:scenario}
\end{figure*}

In this paper, we propose an off-axis jet scenario to explain the observed properties of GRB~190829A.
If the jet is viewed off-axis, 
the relativistic beaming and Doppler effects 
cause the prompt emission to be dimmer and softer than the on-axis viewing case
\citep{ioka2001,ioka2018,yamazaki2002,yamazaki2003a,yamazaki2004a,yamazaki2004b,yamazaki2003b,Salafia2015,Salafia2016}.
Some low-luminosity GRBs may be explained by this context
\citep[e.g.,][]{yamazaki2003b,Ramirez-Ruiz2005},
although there are some counterarguments
\citep[e.g.,][]{Matsumoto2019}.
This model may also explain observed achromatic behavior of early X-ray and optical/IR afterglow with a maximum 
at $1.4\times10^3$~s.
For the off-axis afterglow \citep[e.g.,][]{granot2002}, the bulk Lorentz factor of the jet is initially so high that the afterglow emission is very dim because of the relativistic beaming effect.
As the jet decelerates, the beaming effect becomes weak, resulting in the emergence of a rising part in afterglow light curves.
After the peak of the emission, the jet has smaller Lorentz factors so that the light curve only weakly depends on the viewing angle.
This paper is organized as follows.
In \S~2, we construct our afterglow model following \citet{Huang2000}.
For simplicity, the jet is assumed to be uniform, and structured jets \citep[e.g.,][]{Rossi2002,Zhang2002,Zhang2004} are not considered.
In \S~3, we show that our model explains the observed afterglow in the X-ray, optical, and radio bands.
In order to explain the observed data, we need two jets with narrow and wide opening angles (see Fig.~\ref{fig:scenario}).
The former is viewed off-axis, while the other is not.
Such a two-component jet model \citep[e.g.,][]{Peng2005,Racusin2008} might be supported by the fact that the prompt emission has two independent components.
In \S~4, VHE gamma-ray  flux at $2\times10^4$~s is estimated.
In \S~5, using a simple model, we discuss the on-axis prompt emission properties of the narrow jet.
Section~6 is devoted to a discussion.
In this paper,  cosmological parameters $H_0=71$~km~s$^{-1}$~Mpc$^{-1}$, $\Omega_M=0.27$, and $\Omega_\Lambda=0.23$ \citep{spergel2003} are adopted following \citet{Chand2020}, whose values of $E_{\rm iso,\gamma}$ and $E_p$ are directly used in this paper.
Then, the luminosity distance to GRB~198029A is 0.35~Gpc.

\section{Model description of jetted afterglow}
In this section, following \citet{Huang2000}, we describe a model of jet dynamics and associated synchrotron emission.
Let $t_b$ and $r$ be the time and radial coordinates, respectively, in the rest frame of the central engine located at the origin, $r=0$.
In this frame, the polar angle $\theta$ is set such that the central axis of the jet corresponds to $\theta=0$.
We assume a uniform jet with a thin shell emitting region at radius $R$. 
The jet velocity is 
\begin{equation}
\beta c=\cfrac{dR}{dt_b}~~,
\end{equation}
where $c$ is the speed of light, and the bulk Lorentz factor is $\Gamma=1/\sqrt{1-\beta^2}$. 
In the central engine frame, the jet is ejected from the central engine at $t_b=0$.
Initially, the jet has the opening half-angle $\theta_0$, isotropic-equivalent kinetic energy $E_{\rm iso,K}$ 
and the initial bulk Lorentz factor $\Gamma_0$.
Ambient interstellar matter (ISM) is assumed to be uniform with the number density $n_0$.
The jet decelerates via interactions with ISM and forms a thin shell.
The decrease of  $\Gamma$ is given by 
\begin{equation}
\cfrac{d \Gamma}{dm}=-\cfrac{\Gamma^2-1}{M_{ej}+\epsilon m+2(1-\epsilon)\Gamma m}~~,
\label{eq:evolve_Gamma}
\end{equation}
where $M_{\rm ej}=E_{\rm iso,K}/\Gamma_0c^2$ and $\epsilon$ are ejecta mass and the radiative efficiency,
respectively, and $m$ is 
the swept-up mass \citep[see, e.g.,][]{Huang2000}.
For our parameters adopted in \S~3, the value of $\epsilon$ is too small to affect our results.
It is geometrically related to the shell radius $R$ and the jet opening half-angle $\theta_j$ as
\begin{equation}
\cfrac{dm}{dR}=2\pi R^2(1-\cos\theta_j)n_0m_p~~,
\end{equation}
where $m_p$ is the mass of the proton.
We assume that the jet spreads laterally at the sound speed $c_s$ measured in the shell comoving frame
\citep[see][for more detailed treatments on the dynamical evolution  calculations of the gas temperature]{Peer2012,Nava2013},
and set the increase of the jet opening half-angle as
\begin{equation}
\cfrac{d \theta_j}{d t_b} =\cfrac{c_{s}}{\Gamma R}~~.
\end{equation}
Solving Eqs.~(1)--(4), we get the jet dynamics, that is, $\Gamma$ and $\theta_j$ as a function of time.
Following the standard convention, their time evolution is shown with the on-axis observer ($\theta=0$) time $t$ which is related to $t_b$ by
\begin{equation}
\cfrac{dt}{dt_b}=1-\beta ~~.
\end{equation}

In calculating synchrotron radiation, we assume that microphysics parameters $\epsilon_e$ and $\epsilon_B$, the energy fractions of internal energy going into radiating electrons and magnetic field, are constant.
The electron energy distribution in the emitting thin shell has a power-law form with index $p$.
In the slow cooling regime, the electron spectrum has a break at the electron cooling Lorentz factor $\gamma_c$,
where we take into account the SSC cooling in the Thomson limit as well as synchrotron energy losses 
\citep{Dermer2000,Sari2001,Zhang2001ssc}. 
Then, it has a form $N(\gamma_e)\propto\gamma_e{}^{-p}$ when $\gamma_m<\gamma_e<\gamma_c$
and $N(\gamma_e)\propto\gamma_e{}^{-p-1}$ when $\gamma_c<\gamma_e$. 

We assume that the observer's line of sight is $\theta=\theta_v$. The flux density $F_{\nu}$ 
of the afterglow emission that arrives at the observer time $T$ is  obtained by integrating the emissivity over the equal arrival time surface determined by
\begin{equation}
\int \frac{1-\beta \cos \Theta}{\beta c} dR= \frac{T}{1+z}~~,
\end{equation}
where $\Theta$ is the angle between the radial direction at each emitter position and the line of sight
\citep[e.g.,][]{Granot1999}.
In summary, parameters of the present model are 
isotropic-equivalent kinetic energy $E_{{\rm iso,K}}$,
initial Lorentz factor $\Gamma_0$, initial jet opening half-angle $\theta_0$, ISM density $n_0$, microphysics parameters $\epsilon_e$ and $\epsilon_B$, electron power-law index $p$, and the viewing angle $\theta_v$.


\section{Numerical results for afterglow emission}
In this section, we show our numerical results of synchrotron afterglow emission in the X-ray ($10^{18}$~Hz), optical (V-band), and radio (1.3 and  15.5~GHz) bands, and compare them with observation data of GRB~190829A.
The X-ray data are extracted from the {\it Swift} team website\footnote{https://www.swift.ac.uk/xrt\_curves/00922968/} \citep{Evans2007,Evans2009} which provides us with
the integrated energy flux in the 0.3--10~keV band and
the photon indices at some epoch. 
The index was around 2.2 at any time. On the other hand, we numerically calculate the energy flux density $F_{\nu=10^{18}{\rm Hz}}$.
In order to compare theoretical and observational results, we convert the observed integrated energy flux to the flux density at $10^{18}$~Hz assuming that the photon index is 2.2 at any time.
The optical V-band data (before the absorption correction) are obtained from \citet{Chand2020}.
In our numerical calculation, we take the V-band extinction $A_{\rm V} = 1.5$~mag \citep{Chand2020}.
The radio data are taken from \citet{Rhodes2020}.


\subsection{Off-axis afterglow emission from a narrow jet}
\label{subsec:single}

\begin{figure*}\vspace*{10pt}
\begin{minipage}{0.45\linewidth}
\centering
\includegraphics[width=1.0\textwidth]{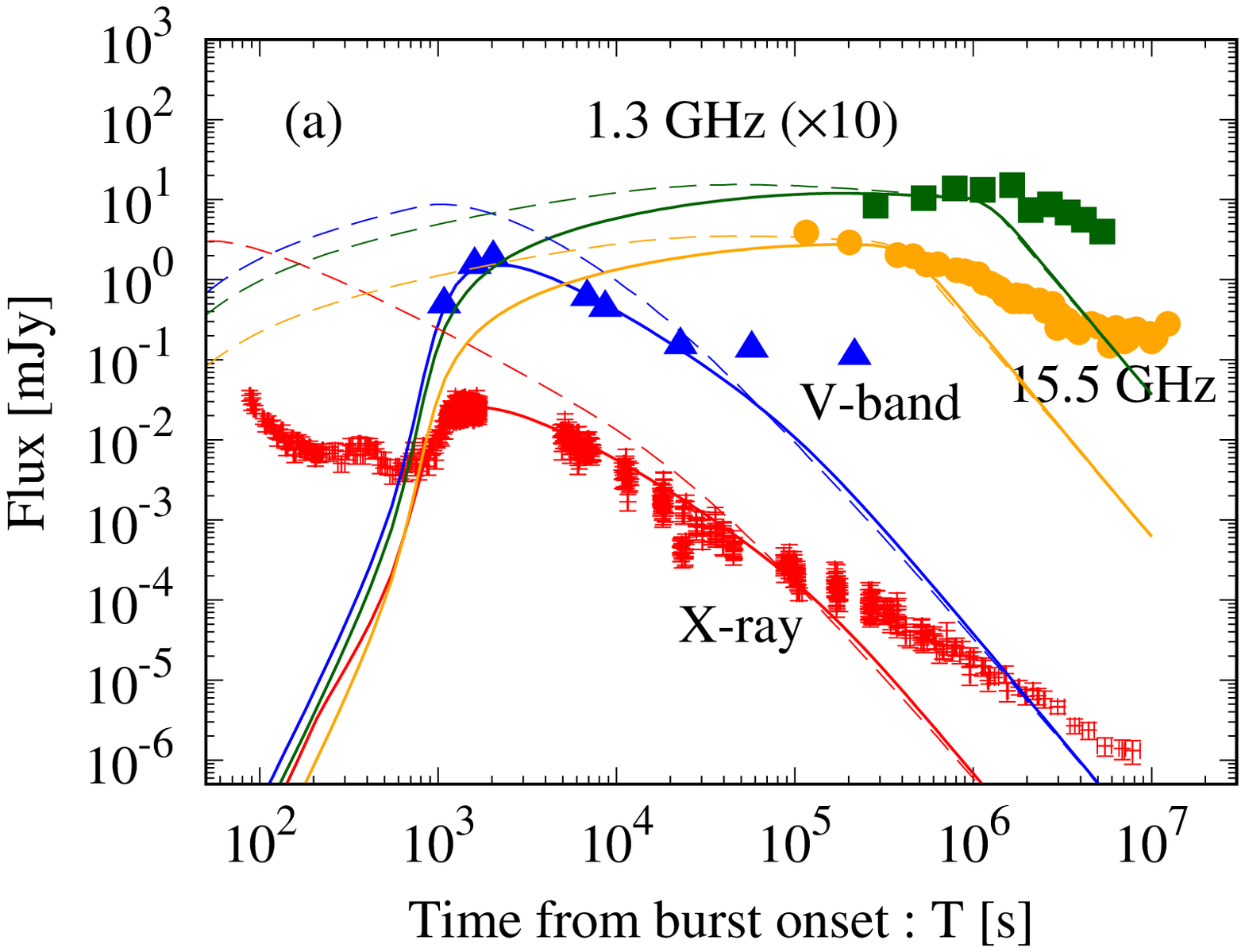}
\end{minipage}
\begin{minipage}{0.45\linewidth}
\centering
\includegraphics[width=1.0\textwidth]{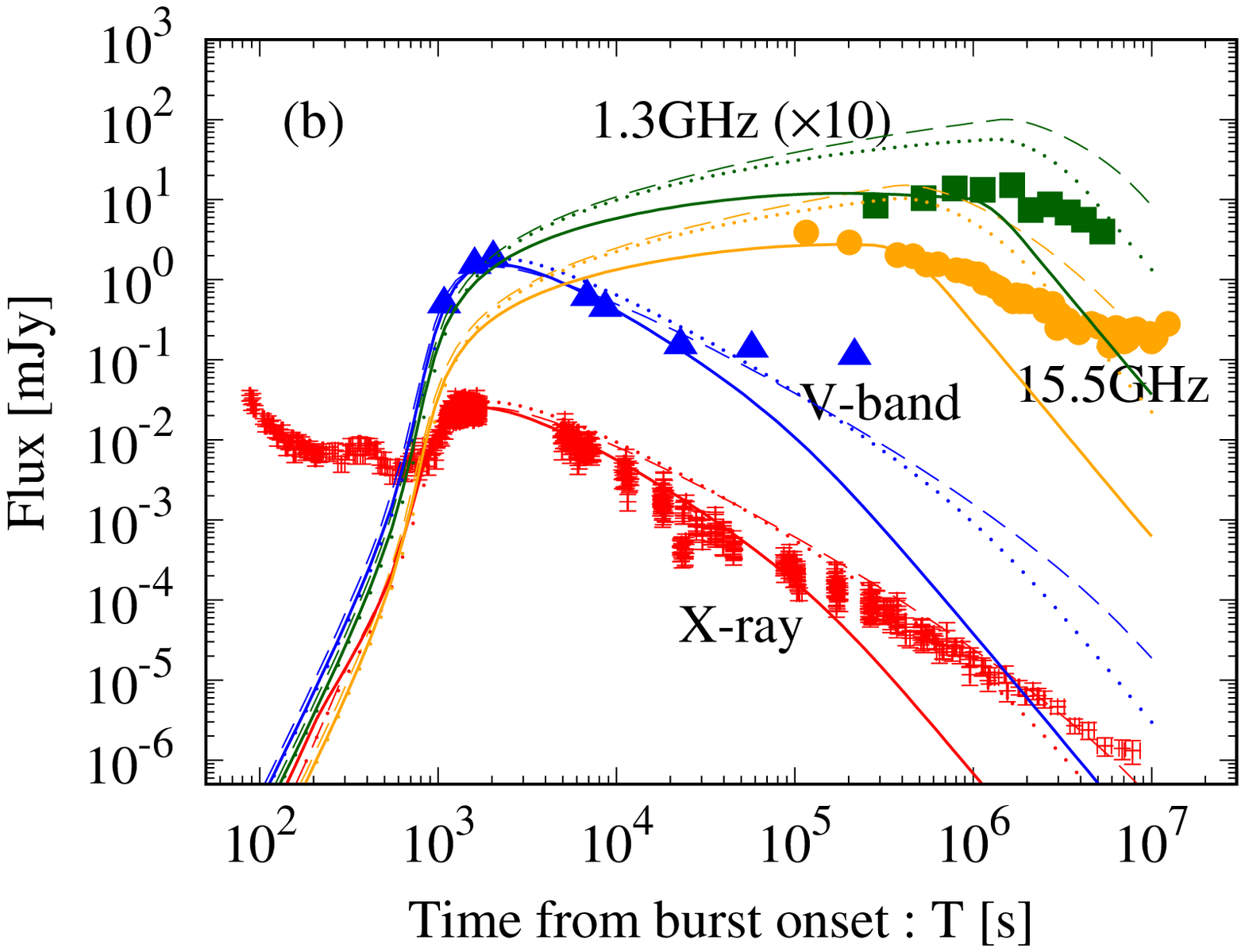}
\end{minipage}
\begin{minipage}{0.45\linewidth}
\centering
\includegraphics[width=1.0\textwidth]{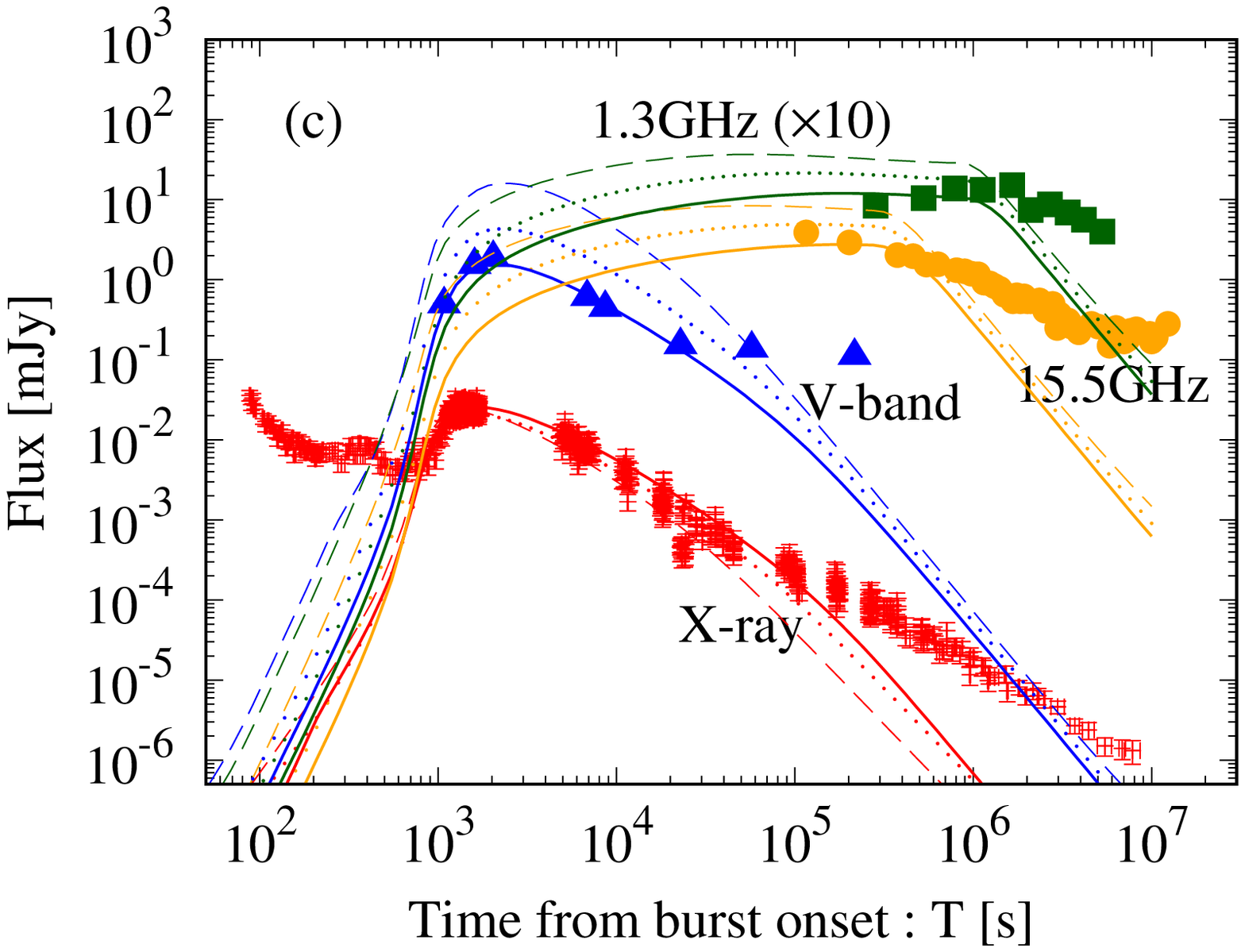}
\end{minipage}
\caption{
Afterglow light curves 
from single jets with various parameter sets 
in the X-ray~($10^{18}$~Hz: red), optical (V-band: blue) and radio bands (1.3~GHz: orange, 15.5~GHz: green), which is compared with the observed data of GRB~190829A (X-ray: red points, V-band: blue triangles, 1.3~GHz: orange filled-circles, 15.5~GHz: green squares). 
Here we choose model parameters to explain observed X-ray peak at about $1.4\times10^4$~s.
In all panels (a), (b) and (c), solid lines are calculation results for our fiducial parameter set for ``narrow jet'' emission
($\theta_0=0.015$~rad, $\theta_v=0.031$, $E_{\rm iso,K}=4.0\times10^{53}$~erg, $\Gamma_0=350$, $n_0 =0.01$~cm$^{-3}$, $\epsilon_e=0.2$, $\epsilon_B=5.0\times10^{-5}$ and $p=2.44$).
In panel~(a), dashed lines show the results for on-axis viewing case ($\theta_v=0$ with other parameters unchanged).
In panel~(b), dashed and dotted lines are for cases of wider jets $\theta_0=0.1$ and 0.05~rad, respectively, to explain the observed X-ray afteglow
in late phase ($10^5{\rm s}\la T\la10^7{\rm s}$). 
We take parameters  $\theta_v=0.116$~rad, $\epsilon_e=0.19$ and $\epsilon_B=1.0\times10^{-5}$ for the former,
and $\theta_v=0.066$~rad and $\epsilon_B=2.0\times10^{-5}$ for the latter, with the other parameters being fiducial.
In panel~(c), dashed and dotted lines are for cases of higher ambient density $n_0 =1.0$ and 0.1~cm$^{-3}$, respectively, keeping
the narrow initial jet opening angle ($\theta_0=0.015$~rad).
Adopted parameters are $\theta_v=0.035$~rad, $E_{\rm iso,K}=2.0\times10^{53}$~erg,  $\epsilon_e=0.4$ and $\epsilon_B=1.0\times10^{-5}$ for the former,
while $\theta_v=0.032$~rad, $E_{\rm iso,K}=1.0\times10^{53}$~erg, $\epsilon_e=0.3$ and $\epsilon_B=1.0\times10^{-5}$ for the latter
(the other parameters are unchanged from the fiducial case).
In the cases of (b) and (c), optical and/or radio data cannot be explained.
}
\label{lightcurve_n}
\end{figure*}

\begin{figure}\vspace*{10pt}
\includegraphics[width=0.45\textwidth]{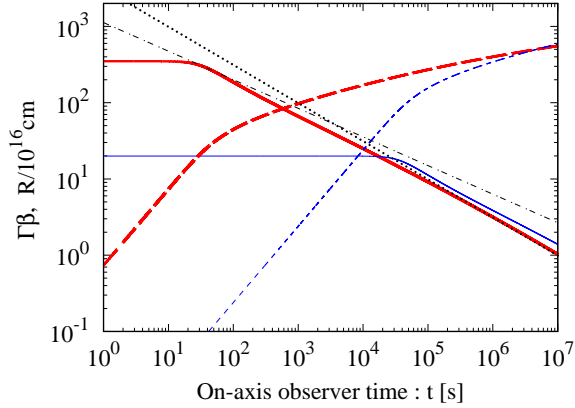}
\caption{
The four-velocity $\Gamma\beta$ (solid lines) and radius $R$ (dashed lines) of narrow (thick-red lines) and wide (thin-blue lines) jets 
with fiducial parameters
as a function of the on-axis observer time $t$.
The black dot-dashed and dotted lines represent analytical scalings $\Gamma\beta\propto t^{-3/8}$ (adiabatic evolution without sideway expansion) 
and $\Gamma\beta\propto t^{-1/2}$ (adiabatic evolution with sideway expansion), respectively. 
}
\label{gamma-radius}
\end{figure}

\begin{figure}\vspace*{10pt}
\includegraphics[width=0.45\textwidth]{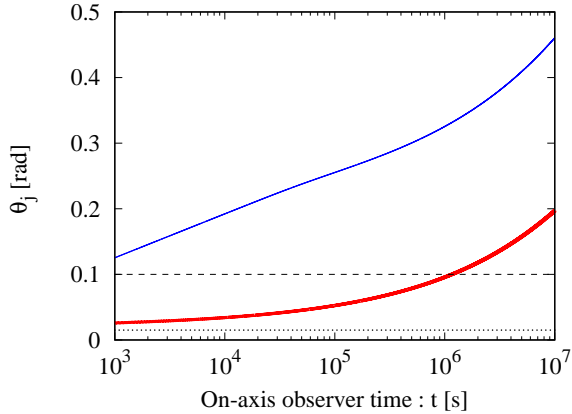}
\caption{
Jet opening half-angle $\theta_j$ as a function of the on-axis observer time $t$.
Thick-red-solid and thin-blue-solid curves are for narrow  and wide jets
with fiducial parameters,
respectively. Two horizontal dotted and dashed lines are the initial values (0.015~rad and 0.1~rad for
the narrow and wide jets, respectively).
}
\label{theta}
\end{figure}

Here we consider a single jet viewed off-axis in order to discuss the observed X-ray and optical bumps around $T\sim1.4\times10^2$~s.
We adopt $\theta_v=0.031$~rad
($1.7^\circ$),
$\theta_0=0.015$~rad~($0.86^\circ$),  $E_{\rm iso,K}=4.0\times10^{53}$~erg, $\Gamma_0=350$,  $n_0 =0.01$~cm$^{-3}$, $\epsilon_e=0.2$, $\epsilon_B=5.0\times10^{-5}$ and $p=2.44$ as a fiducial parameter set.
The initial opening half-angle is small, so that we refer to ``narrow jet''  in the following.
However, the jet is still ``fat'' in the sense $\theta_0>\Gamma_0^{-1}$, so that the jet dynamics is able to be discussed as in a standard manner.
Solid lines in Fig.~\ref{lightcurve_n}(a)-(c)
show the results for our fiducial parameters.
Our off-axis afterglow model well explains the observational results of early X-ray and optical afterglow from about $8\times10^2$ to $2\times10^4$~s.
An achromatic behavior in the X-ray and optical bands is evident.
The off-axis afterglow starts with a rising part because of the relativistic beaming effect \citep{granot2002}.
As the jet decelerates, the observed flux increases.
When the jet Lorentz factor becomes $\Gamma\sim(\theta_v - \theta_0)^{-1}=65$, the afterglow light curve takes a maximum.
After that, the observed flux is almost the same as that in the case of on-axis viewing 
($\theta_v=0$: dashed lines in Fig.~\ref{lightcurve_n}(a)).
If we assume the adiabatic evolution ($\Gamma\propto t^{-3/8}$), the observer time of the flux maximum
is analytically given by
\begin{equation}
T_{pk}\sim(1+z)\left(\cfrac{3E_{{\rm iso,K}}}{4\pi n_0m_pc^5}\right)^{\frac{1}{3}}(\theta_v-\theta_0)^{\frac{8}{3}}~~.
\label{eq:jetbreaktime}
\end{equation}
For our 
fiducial
model parameters, we get $T_{pk}\sim2\times10^3$~s,
which is consistent with our numerical results within a factor of two.
As shown in Fig.~\ref{gamma-radius} (thick-red-solid and dot-dashed lines), the scaling $\Gamma\propto t^{-3/8}$ is roughly a good approximation until $t\sim10^4$~s, since the jet spreading effect is not significant (see the thick red line in Fig.~\ref{theta}) and the radiative efficiency $\epsilon$ in Eq.~(\ref{eq:evolve_Gamma}) is small
($\epsilon\la0.04$ for $t\ga10^2$~s).
Hence,  the estimate of $T_{pk}$ by Eq.~(\ref{eq:jetbreaktime}) on the assumption of the spherical adiabatic expansion is justified.
For comparison, the dashed lines in 
 Fig.~\ref{lightcurve_n}(a)
show the light curves in the on-axis viewing case ($\theta_v=0$), in which the X-ray flux peaks much earlier than the optical one \citep{sari1998}.
 
In the fiducial case,
after several tens of thousand seconds after the burst onset, our numerically calculated X-ray light curve deviates from the observed data 
as shown by red solid lines in Fig.~\ref{lightcurve_n}(a)-(c).
Before that epoch, the sideway expansion of the jet is not significant (thick-red-solid line in Fig.~\ref{theta}).
As the jet decelerates, the jet becomes trans-relativistic ($\Gamma\la10$) around $t\sim10^5$~s, and then $\theta_j$ begins rapid increase \citep{Zhang2009}\footnote{
In the past literature, it used to be assumed that a relativistic jet rapidly decelerates and its opening angle increases exponentially just after the jet break time which is given by 
$t_{\rm jet}\sim(3E_{\rm iso,K}/4\pi n_0m_pc^5)^{1/3}\theta_0^{8/3}$
\citep{sari1999},
and for our 
fiducial
model parameters, we get $t_{\rm jet}\sim2\times10^3$~s.
However, as shown by relativistic hydrodynamics  simulation by \citet{Zhang2009}, the lateral expansion is not significant until the trans-relativistic phase.
}.
Then, our numerical result shows that the jet dynamics asymptotically reaches  the scaling $\Gamma\propto t^{-1/2}$ (black-dotted line in Fig.~\ref{gamma-radius}),
at which the observed X-ray flux follows the scaling $F_\nu\propto t^{-p}=t^{-2.44}$ \citep{sari1999}.
This slope is much steeper than observed.
Hence, it is  difficult for the narrow jet 
with fiducial parameters
to explain the observed late X-ray afterglow as well as
radio emission at late times.

Since the fiducial value of the initial jet opening half-angle $\theta_0$  is small, 
 the predicted X-ray afterglow emission is dim at the late epoch ($T\ga10^5$~s).
We examine whether a single jet could explain the X-ray afterglow till $T\sim10^7$~s with larger $\theta_0$ (see Fig.~\ref{lightcurve_n}(b)).
If we take $\theta_0=0.1$~rad (dashed line in Fig.~\ref{lightcurve_n}b) and 0.05~rad (dotted lines in fig.~\ref{lightcurve_n}b),
with the viewing angle $\theta_v=0.116$ and 0.066~rad, respectively, to have similar $T_{pk}$ given by Eq.~(\ref{eq:jetbreaktime}),
then we may explain the achromatic peaks in the X-ray and optical bands at $1.4\times10^3$~s and  
the late X-ray emission at the same time (for the other adopted parameters, see caption in Fig.~\ref{lightcurve_n}).
In these cases, however, numerically calculated radio emission  is brighter than observed.
This is simply because the number of radio emitting electrons increases with the jet solid angle.
Hence, such large-$\theta_0$ cases are disfavoured.

The ISM  density is  low in the fiducial case ($n_0 =0.01$~cm$^{-3}$).
We discuss whether a larger $n_0$ could explain the observational results or not.
The dashed and dotted lines in Fig.~\ref{lightcurve_n}(c) show the results
for $n_0 =1.0$~cm$^{-3}$ 
and 0.1~cm$^{-3}$, respectively (for the other model parameters, see caption in Fig.~\ref{lightcurve_n}).
In these cases, the observed early X-ray emission is well explained.
However, the predicted optical flux around the peak time and  the radio emission at $1\times10^5{\rm s}\la T\la4\times10^5$~s become brighter than observed.
Hence, the case of larger $n_0$ is unlikely.

When our off-axis jet model explains the achromatic peaks in the X-ray and optical bands at 1400~s, it is hard for a single jet to describe the emission from all wavelengths at any time.
It is necessary to consider the case where the narrow jet  propagates into rarefied  medium 
as given  by the fiducial parameters.
In the next section \S~3.2, we will add another component to enhance the late X-ray and radio fluxes.


\begin{figure}\vspace*{10pt}
\includegraphics[width=0.45\textwidth]{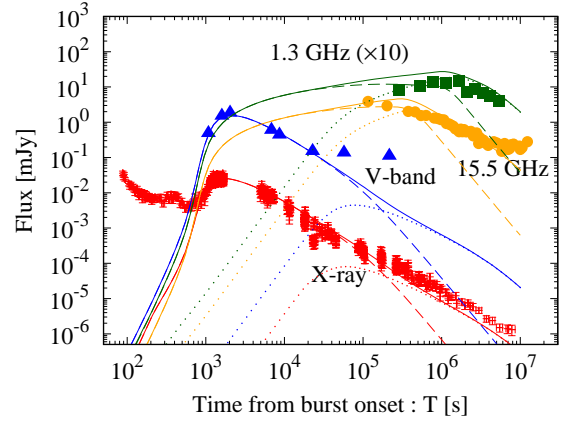}
\caption{
Afterglow light curves calculated by our two-component jet model
--- solid lines are the sum of the narrow (dashed lines) and wide (dotted lines) jets with fiducial parameters.
The solid lines describe the best-fitted model in this paper.
The meanings of colors and observed data are the same as in Fig.~\ref{lightcurve_n}.
Note that late-time ($t\ga5\times10^4$~s) optical band is dominated by supernova component  \citep{Hu2020}.
}
\label{lightcurve}
\end{figure}


\begin{figure*}\vspace*{10pt}
\begin{minipage}{0.45\linewidth}
\centering
\includegraphics[width=1.0\textwidth]{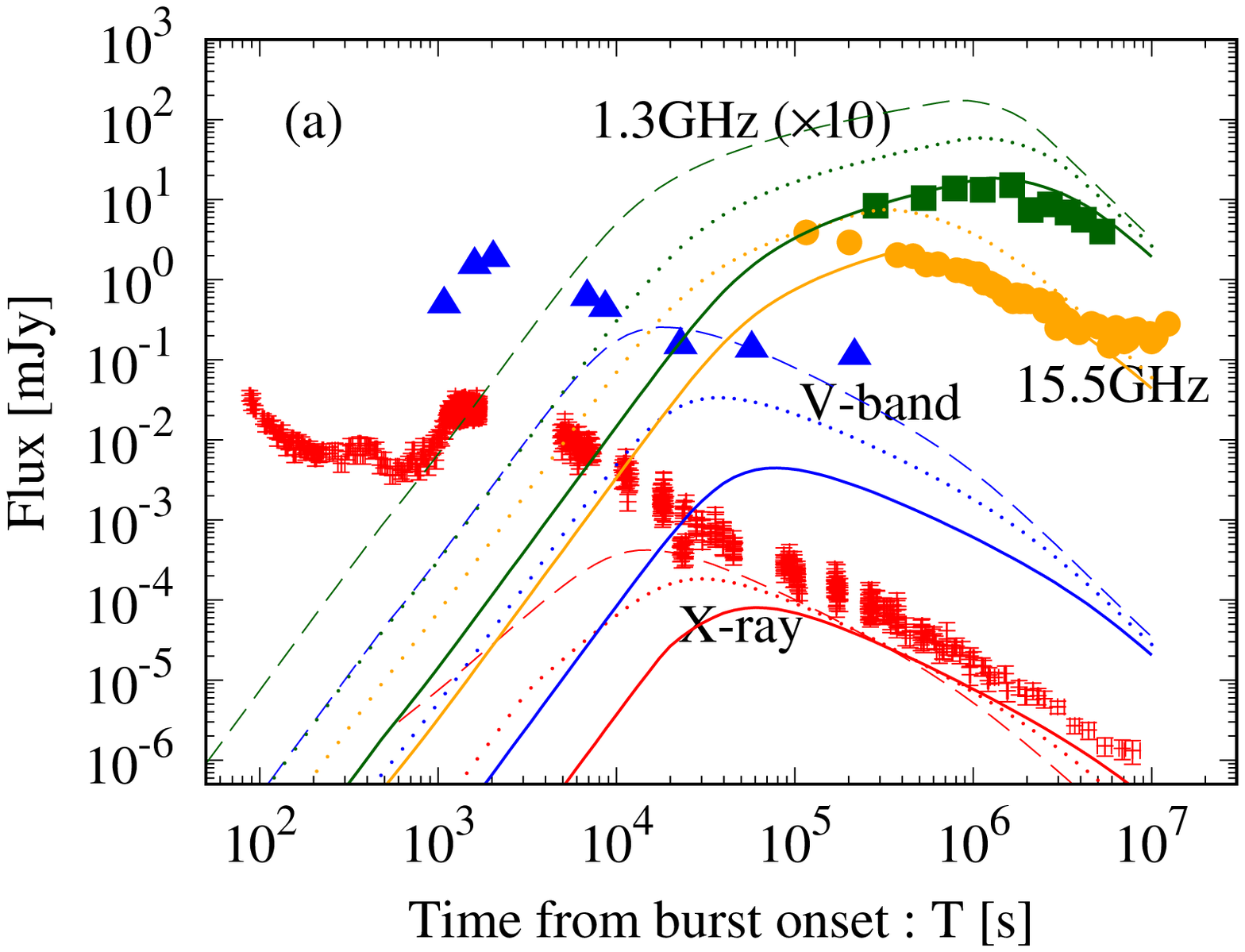}
\end{minipage}
\begin{minipage}{0.45\linewidth}
\centering
\includegraphics[width=1.0\textwidth]{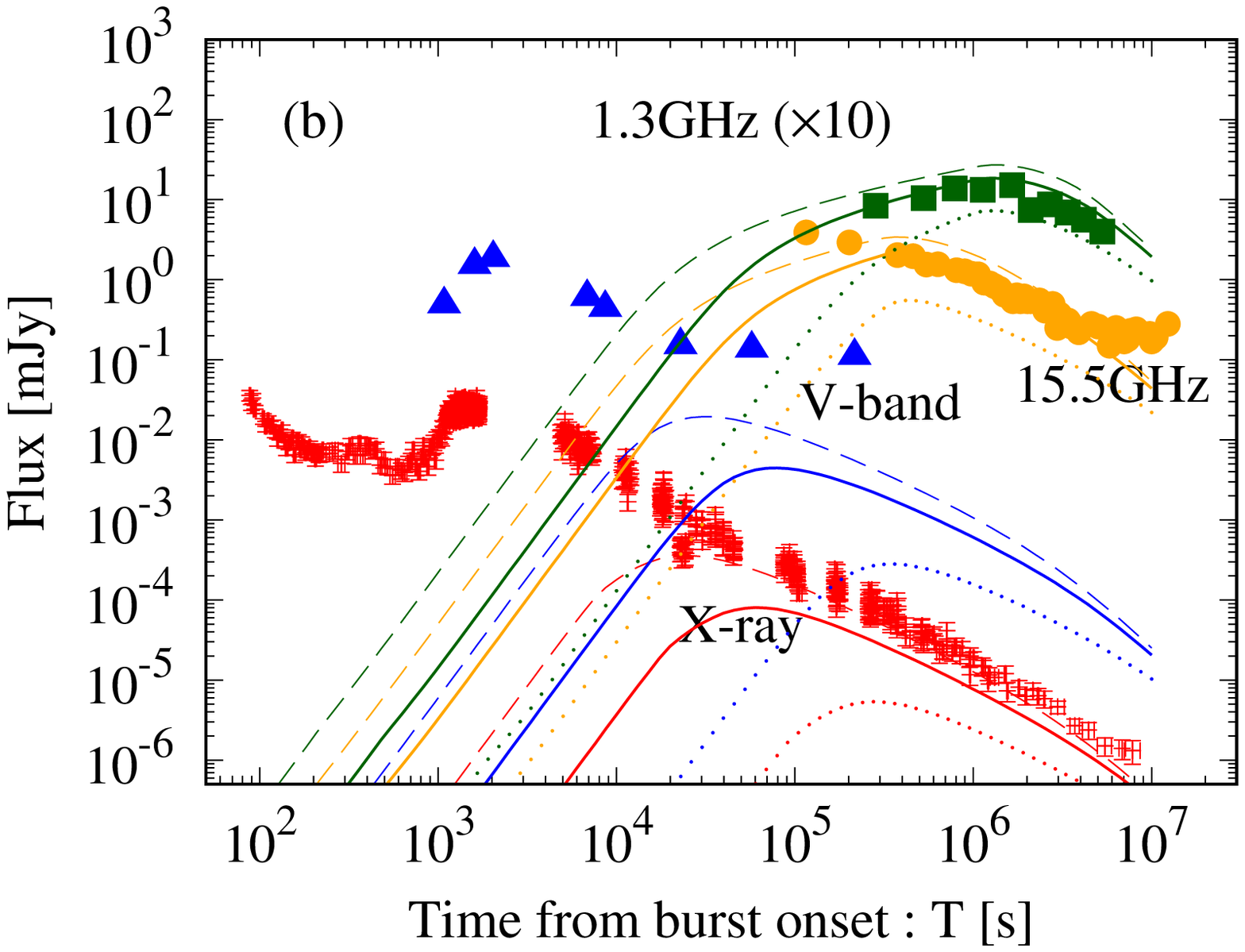}
\end{minipage}
\begin{minipage}{0.45\linewidth}
\centering
\includegraphics[width=1.0\textwidth]{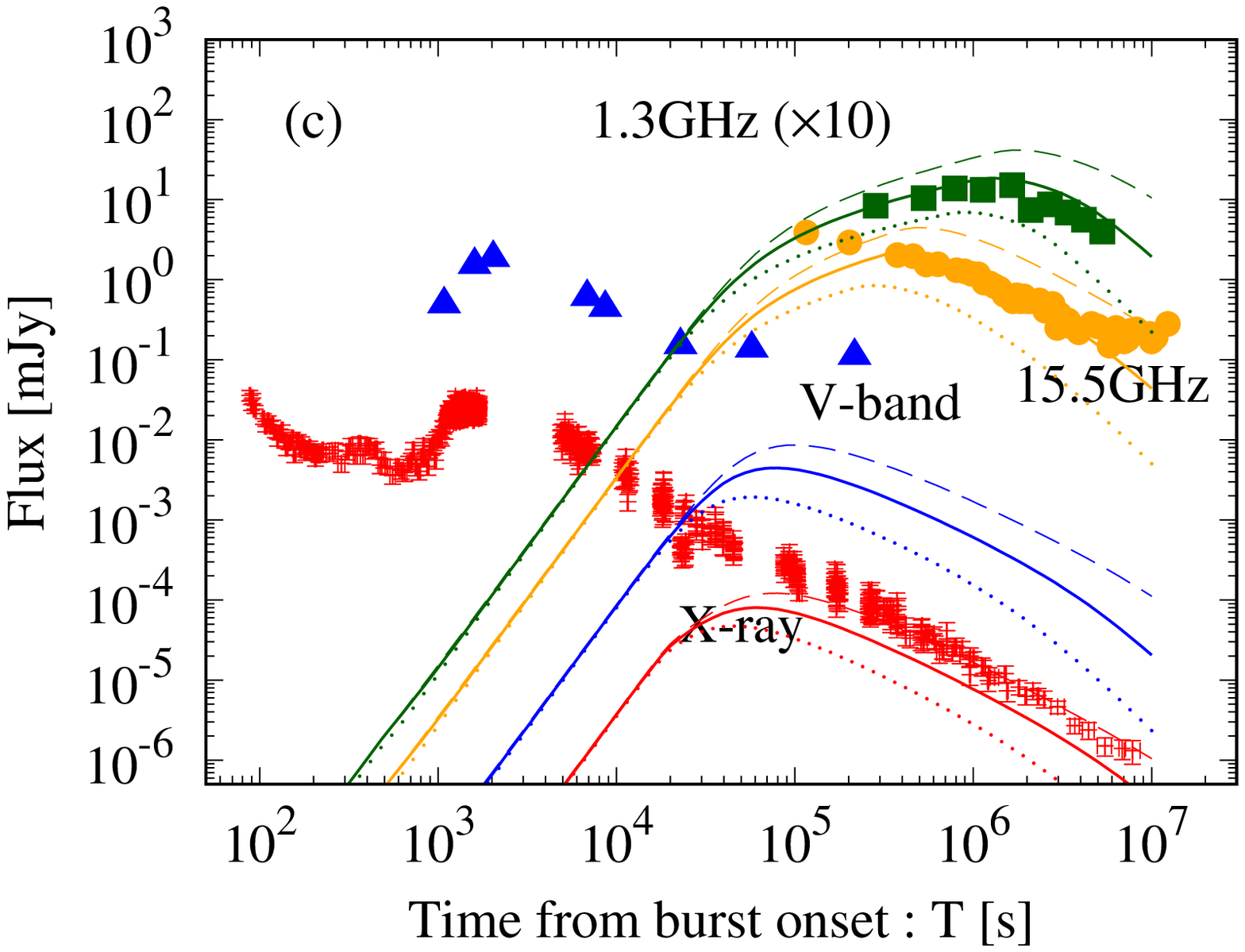}
\end{minipage}
\caption{
Parameter dependence of afterglow emission from the ``wide jet''  explaining the late ($T\ga10^5$~s) X-ray and radio data.
The meanings of colors and observed data are the same as in Fig.~\ref{lightcurve_n}.
Solid lines in all panels (a), (b) and (c) show results for the fiducial parameter set 
($\theta_v=0.031$~rad, $\theta_0=0.1$~rad, $E_{\rm iso,K}=2.0\times10^{53}$~erg, $\Gamma_0=20$, $n_0 =0.01$~cm$^{-3}$, $\epsilon_e=0.4$, 
$\epsilon_B=1.0\times10^{-5}$ and $p=2.2$), which correspond to dotted lines in Fig.~\ref{lightcurve}.
In panel~(a), ISM density $n_0$ is changed --- dashed and dotted lines are for $n_0 =1$ and 0.1~cm$^{-3}$, respectively, with the other parameters unchanged.
In panel~(b), only the initial Lorentz factor of the jet $\Gamma_0$ is varied (dashed line: $\Gamma_0=30$, dotted line: $\Gamma_0=10$).
In panel~(c), we show the results for different initial jet opening half-angles $\theta_0 =0.2$ (dashed lines) and 0.05~rad (dotted lines) with other parameters being fiducial.
}
\label{lightcurve_w}
\end{figure*}

\subsection{Two-component jet model}
In this subsection, we consider a two-component jet model, in which another ``wide jet'' is introduced in addition to the narrow jet considered in \S~\ref{subsec:single}.
The observed flux is simply the superposition of each jet emission components. The parameters of the narrow jet are the same as those given in \S~\ref{subsec:single}.
For the wide jet, we adopt $\theta_0=0.1$~rad~($5.73^\circ$), $E_{{\rm iso,K}}=2.0\times10^{53}$~erg, $\Gamma_0=20$, $\epsilon_e=0.4$, $\epsilon_B=1.0\times10^{-5}$, and $p=2.2$ as a fiducial parameter set.
The values of $\theta_v$ and $n_0$ are common for both jets.
It is assumed that the central axes of the two jets are identical ($\theta=0$: see Fig.~\ref{fig:scenario})
and that the two jets depart from the central engine ($r=0$) at the same time.

One can find 
in Figure~\ref{lightcurve}
that early achromatic  peaks in the X-ray and optical bands are explained by the off-axis narrow jet emission (dashed lines in the right panel), and that  the late X-ray 
and radio afterglow is interpreted with the wide jet emission (dotted lines).
Our model systematically underpredicts the late X-ray emission, however, considering our model uncertainties
this may not be a serious problem (see more detailed discussion in \S~6).
As shown in the thin-blue-solid line in Fig.~\ref{gamma-radius}, the wide jet does not decelerate until $t\sim5\times10^4$~s, since it is heavy  with a low bulk Lorentz factor, $\Gamma_0=20$. 
Our numerical result 
(dotted lines in Fig.~\ref{lightcurve})
shows that X-ray and optical flux becomes maximum around this epoch.
The wide jet becomes trans-relativistic ($\Gamma\la10$) at $t\sim10^5$~s and finally enters to the Newtonian phase at $t\ga10^{6}$~s.
We find that for $10^5$~s~$\la t\la10^7$~s,  the absorption frequency $\nu_a$, typical frequency $\nu_m$ and cooling frequency $\nu_c$ obeys the relation $\nu_a<\nu_m<\nu_c$.
The value of $\nu_c$ is located between the optical and the X-ray bands, and  $\nu_m$ is lower than the optical band.
Then, the X-ray and optical fluxes follow the scalings  $F_\nu\propto t^{-(3p-4)/2}=t^{-1.3}$ and $F_\nu\propto t^{-3(5p-7)/10}=t^{-1.2}$, respectively \citep[e.g.,][]{Gao2013}, which is consistent with the observation.
The typical frequency $\nu_m$ decreases with time, and
at $t\sim4\times10^5$~s, it crosses 15.5~GHz, at which the 15.5~GHz light curve has a peak.
After that, the flux follows the scaling $F_\nu\propto t^{-3(5p-7)/10}=t^{-1.2}$ \citep[e.g.,][]{Gao2013}.
Subsequently, $\nu_m$ intersects 1.3~GHz at  $t\sim1\times10^6$~s, and the 1.3~GHz flux takes maximum, after which the flux decays in the same manner.
These radio behavior is roughly consistent with the observation.

We consider parameter dependence for the wide jet emission.
Since the parameters for the narrow jet are already determined, here we fix the viewing angle ($\theta_v=0.031$~rad).
First, the ISM density $n_0$ is changed with the other parameters being fiducial.
 The solid, dotted and dashed lines in Fig.~\ref{lightcurve_w}(a) are for $n_0 =0.01$ (fiducial), 0.1 and 1.0~cm$^{-3}$, respectively.
Our numerically calculated optical and radio emissions  are brighter than observed for large $n_0$.
The large-$n_0$ case is again disfavoured also for the wide jet as already seen for the narrow jet case.
In the framework of our two-component jet model, the low density ($n_0\sim0.01$~cm$^{-3}$) is necessary.
Second, we alter
only the initial Lorentz factor  $\Gamma_0$  (Fig.~\ref{lightcurve_w}(b)).
Our numerical result for $\Gamma_0=30$ (dashed lines in Fig.~\ref{lightcurve_w}(b)) is brighter than the optical observed data.
Also, the total flux of the narrow  and wide jets exceeds the observed X-ray data.
For  $\Gamma_0=10$, the calculated radio and X-ray fluxes (dotted lines in Fig.~\ref{lightcurve_w}(b)) are much dimmer than observed.
Even if other parameters are changed to match the observed X-ray light curve, 
the predicted radio emission becomes  brighter than  observed.
Hence, the best value for $\Gamma_0$ is about 20.
Third, we consider the cases of different values of $\theta_0$ as shown in Fig.~\ref{lightcurve_w}(c) to describe the late X-ray emission.
If $\theta_0$ is larger ($\theta_0=0.2$~rad), the calculated X-ray light curve  fits observation result.
However, the radio emissions are  brighter than  observed.
On the other hand, if $\theta_0$ is smaller ($\theta_0=0.05$~rad), then the radio flux becomes dim, while the X-ray observation  is difficult to be explained.
Therefore, the best value for $\theta_0$ of the wide jet is about 0.1~rad.

\section{VHE gamma-ray emission at $2\times10^4$~seconds}

In this section, we estimate the VHE gamma-ray flux at  $2\times10^4$~s along with our two-component jet model considered in \S~3.
For simplicity, the SSC flux at $h\nu=0.1$~TeV is calculated in the Thomson limit \citep{Sari2001}.
The flux attenuation by extragalactic background light at 0.1~TeV is negligible because the source is nearby \citep{Zhang2020b}.

For both narrow and wide jets, we get the bulk Lorentz factor $\Gamma$ of the jets and the synchrotron spectrum $F_\nu$ at $2\times10^4$~s as seed photons for SSC emission.
First, we consider the narrow jet
with fiducial parameters, 
which has the bulk Lorentz factor $\Gamma\simeq20$, the post-shock magnetic field $B\simeq5.4\times10^{-3}$~G, the minimum electron Lorentz factor $\gamma_m\simeq2.2\times10^3$, the electron cooling Lorentz factor 
$\gamma_c\simeq1.3\times10^6~[15/(1+Y)]$
, the typical frequency $\nu_m\simeq7.7\times10^{12}$~Hz, the cooling frequency 
$\nu_c\simeq6.9\times10^{17}~{[15/(1+Y)]}^{2}$~Hz
and the peak flux $F_{\rm max}=F_{\nu=\nu_m}\simeq1.4\times10$~mJy,
where $Y\simeq14$ is the Compton Y parameter.
Then, the break frequencies for the SSC emission \citep{Sari2001} are given by
$\nu_{m}^{\mathrm{IC}}\approx2 \gamma_{m}^{2} \nu_{m}\simeq7.5\times10^{19}$Hz 
and 
$\nu_{c}^{\mathrm{IC}}\approx2 \gamma_{c}^{2} \nu_{c}\simeq2.3\times10^{30}~{[15/(1+Y)]}^{4}$~Hz.
One can find that the observation photon energy $h\nu=0.1$~TeV satisfies $\sqrt{\nu_{m}^{\mathrm{IC}} \nu_{c}^{\mathrm{IC}}}<\nu<\nu_{c}^{\mathrm{IC}}$,
so that the SSC flux is calculated by
\begin{eqnarray}
F_{\nu}^{\mathrm{SSC}} &\approx &0.5R \sigma_{T} n_0 F_{\rm max } 
\cfrac{(p-1)}{(p+1)}\left(\cfrac{\nu}{\nu_{m}^{\mathrm{IC}}}\right)^{(1-p) / 2} \nonumber \\ 
&&\times \left[2 \cfrac{(2 p+3)}{(p+2)}-\cfrac{2}{(p+1)(p+2)}+\ln \left(\cfrac{\nu_{c}^{\mathrm{IC}}}{\nu}\right)\right]\ ~~, \label{eq1} 
\end{eqnarray}
where $\sigma_{T}$ is the Thomson cross section.
Hence, the SSC energy flux from the narrow jet is estimated as $\nu F_\nu^{\rm SSC}\sim1.2\times10^{-11}$ erg~s$^{-1}$cm$^{-2}$.
Since the jet energy is large, we have a lot of seed photons from its own synchrotron radiation to get detectable SSC emission. 
In reality, the Klein-Nishina effect becomes important below $\nu_{c}^{\rm IC}$. The Y parameter at $\gamma_c$ is significantly reduced due to the Klein-Nishina effect, so the VHE gamma-ray flux is expected to have a peak around TeV energies below $\nu_c^{\rm IC}$ in the Thomson limit. Correspondingly, the value of $\nu_c$ would be underestimated.    

Similarly, we calculate the SSC flux from the wide jet
with fiducial parameters.
At $2\times10^4$~s, we have $\Gamma\simeq20$, $B\simeq2.4\times10^{-3}$~G, 
$Y\simeq10$,
$\gamma_m\simeq2.4\times10^3$,
$\gamma_c\simeq2.4\times10^6~{[11/(1+Y)]}$
, $\nu_m\simeq1.0\times10^{12}$~Hz,  
$\nu_c\simeq1.0\times10^{18}~{[11/(1+Y)]}^{2}$~Hz
, and
$F_{\rm max}\simeq0.1$~mJy, so that we obtain
$\nu_{m}^{\mathrm{IC}}\simeq1.1\times10^{19}$~Hz and
$\nu_{c}^{\mathrm{IC}}\simeq1.1\times10^{31}~~{[11/(1+Y)]}^{4}$~Hz
, resulting in $\nu F_\nu^{\rm SSC}\sim2\times10^{-14}$ erg~s$^{-1}$cm$^{-2}$ at $h\nu=0.1$~TeV.
The narrow jet is predominant at VHE gamma-ray band at about $2\times10^4$~s. Just around this epoch, the wide jet is in the transition from the free expansion to the adiabatic deceleration phase.
Indeed, as shown by the dotted lines in the right panel of Fig.~\ref{lightcurve}, we see a rising part in X-ray and optical bands.
After $2\times10^4$~s, the SSC flux from the wide jet might increase just for a while.
However, it would soon start to decrease and keep subdominant.

We discuss the time dependence of the VHE gamma-ray flux between $\sim10^4$ and $\sim10^5$~s in the Thomson limit. 
In this epoch, the observed flux for $\theta_v=0.031$ hardly changes from that for $\theta_v=0$ 
(for narrow jets with fiducial parameters, see Fig.~\ref{lightcurve_n}(a)), 
so that the standard analytical calculation for on-axis observer is a good approximation for the present case.
We find that for both narrow and wide jets with fiducial parameters, 
the break frequencies and observation frequency $h\nu=0.1$~TeV satisfies 
$\nu_{m}^{\mathrm{IC}}<\nu<\nu_{c}^{\mathrm{IC}}$ at any time, so that the SSC flux is given by $F_{\nu}^{\mathrm{SSC}} \propto R \sigma_{T} n_0 F_{\rm max } (\nu/\nu_{m}^{\mathrm{IC}})^{(1-p) / 2}$.
For the narrow jet, the Lorentz factor is approximated to follow the scaling $\Gamma\propto t^{-1/2}$ (see the thick-red-dashed line in Fig.~\ref{gamma-radius}).
Then for synchrotron component, we have $\gamma_m\propto t^{-1/2}$, $\nu_m\propto t^{-2}$ and $F_{\rm max}\propto t^{-1}$ \citep{sari1999}, so that $\nu_{m}^{\mathrm{IC}}\propto t^{-3}$ and 
$F_{\nu}^{\mathrm{SSC}} \propto t^{-(3p-1)/2}=t^{-3.16}$.
On the other hand, as discussed in the previous paragraph,
the VHE flux from the wide jet has brighten, following the analytical scaling $F_{\nu}^{\mathrm{SSC}} \propto t^4$ 
until $t\sim5\times10^4$~s at the transition from the free expansion to the adiabatic deceleration phase.
After this time, the flux decays.
For the jet dynamics $\Gamma\propto t^{-3/8}$ ($\propto t^{-1/2}$), we get the VHE flux $F_{\nu}^{\mathrm{SSC}} \propto t^{-(9p-11)/8}=t^{-1.1}$ ($\propto t^{-(3p-1)/2}=t^{-2.8}$).
Such a time evolution could have been observed with good statistics by more sensitive detectors like CTA.

\section{Prompt emission properties of narrow jet}

The prompt emission of GRB~190829A had smaller values of the peak energy $E_{\rm p}$ and the isotropic-equivalent gamma-ray energy $E_{\rm iso,\gamma}$ than typical long GRBs.
In this section, we discuss whether 
$E_{\rm p}$ and $E_{\rm iso,\gamma}$ from our narrow jet were typical or not if it would have been viewed on-axis ($\theta_v\approx0$).
We consider a very simple analytical model \citep[e.g.,][]{ioka2001,yamazaki2002,yamazaki2003a} assuming an instantaneous emission of an infinitesimally thin shell moving with the Lorentz factor $\Gamma_0=1/\sqrt{1-\beta_0^2}$.
The jet is uniform, whose intrinsic emission properties do not vary with angle, and has a sharp edge at the opening half-angle $\theta_0$.
Then, the viewing-angle dependence of the peak energy, $E_{p}(\theta_v)$, and isotropic-equivalent gamma-ray energy, $E_{\rm iso,\gamma}(\theta_v)$, can be analytically calculated \citep{Donaghy2006,Graziani2006}, and we obtain 
\begin{eqnarray}
R_1 &=&\cfrac{E_{\rm p}(\theta_{v})}{E_{\rm p}(0)} \nonumber  \\
  &=&\cfrac{2(1-\beta_0)(1-\beta_0\cos\theta_{0})}{2-\beta_0(1+\cos\theta_0)} \nonumber  \\
 && \times \cfrac{f(\beta_0-\cos\theta_v)-f(\beta_0\cos\theta_0-\cos\theta_v)}{g(\beta_0-\cos\theta_v)-g(\beta_0\cos\theta_0-\cos\theta_v)} ~~, 
\end{eqnarray}
\begin{eqnarray}
R_2 &=&\cfrac{E_{\rm iso,\gamma}(\theta_{v})}{E_{\rm iso,\gamma}(0)} \nonumber  \\
  &=&\cfrac{(1-\beta_0)^2(1-\beta_0\cos\theta_{0})^2}{\beta_0(1-\cos\theta_0)[2-\beta_0(1+\cos\theta_0)]} \nonumber  \\
 && \times [f(\beta_0-\cos\theta_v)-f(\beta_0\cos\theta_0-\cos\theta_v)]~~,
\end{eqnarray}
where, functions $f$ and $g$ are given by
\begin{eqnarray}
f(z) =
\cfrac{\Gamma_0^2(2\Gamma_0^2-1)z^3+(3\Gamma_0^2\sin^2\theta_v-1)z+2\cos\theta_v\sin^2\theta_v}{|z^2+\Gamma_0^{-2} {\rm sin^2}\theta_v|^\frac{3}{2}}~~, 
\end{eqnarray}
and
\begin{equation}
g(z)=\cfrac{2\Gamma_0^2z+2\cos\theta_v}{|z^2+\Gamma_0^{-2} {\rm sin^2}\theta_v|^\frac{1}{2}}~~,
\end{equation}
respectively \citep[see also][]{Urata2015}.
For our fiducial parameters of our narrow jet given in \S~3.1
($\Gamma_0=350$, $\theta_0=0.015$~rad and $\theta_v=0.031$~rad),
we get
$R_1=3.2\times10^{-2}$ and
$R_2=1.2\times10^{-4}$.

If the narrow jet emitted  Episode~1 of observed prompt emission (see \S~1), that is,
$E_p(\theta_v=0.031)=120$~keV and
$E_{\rm iso,\gamma}(\theta_v=0.031)=3.2\times10^{49}~{\rm erg}$
\citep{Chand2020},
then on-axis quantities are obtained as
$E_p(0)=E_p(\theta_v)/R_1=3.7$~MeV and
$E_{\rm iso,\gamma}(0)=E_{\rm iso,\gamma}(\theta_v)/R_2=2.7\times10^{53}~{\rm erg}$.
These values are within the range for bursts detected so far
\citep[e.g.,][]{zhao2020a}.
The isotropic equivalent kinetic energy of the narrow jet just after the prompt emission is
$E_{\rm iso,K}=4.0\times10^{53}~{\rm erg}$ (see \S~3.1),
so that the efficiency of the prompt emission is calculated as
$\eta_\gamma=E_{\rm iso,\gamma}(0)/(E_{\rm iso,\gamma}(0)+E_{\rm iso,K})\approx0.4$.
On the other hand, if the narrow jet is responsible for  Episode~2 
(that is, $E_p(\theta_v=0.031)=11$~keV and
$E_{\rm iso,\gamma}(\theta_v=0.031)=1.9\times10^{50}~{\rm erg}$),
we obtain $E_p(0)=340$~keV and
$E_{\rm iso,\gamma}(0)=1.6\times10^{54}~{\rm erg}$,
which are again similar to typical long GRBs.
In this case, the efficiency is $\eta_\gamma\approx0.8$.

The observed prompt emission had two episodes (see \S~1), while in \S~3 we showed two jets are required to
explain the observed afterglow. At present, it is unknown if the two episodes corresponds to the two jets.
If the narrow jet causes Episode~1, then the estimated prompt emission efficiency
$\eta_\gamma$ is almost typical,
however  on-axis $E_{\rm p}(0)$ is located at the highest end of the distribution for long GRBs.
On the other hand, if the narrow jet produced Episode~2, then
on-axis $E_{\rm p}(0)$ is smaller though
$\eta_\gamma$ is somewhat higher (but it is still comparable, and one can say that the value is reasonable
considering very simple approximation of our prompt emission model).
Episode~1 and 2 may be emitted from narrow and wide jets, respectively.  
Note that if the wide jet  emits Episode~2, its efficiency is small,
$\eta_\gamma\approx5\times10^{-3}$,
so that it might be natural that the narrow jet causes both Episode~1 and 2.

In this section, we simply assumed that the prompt emission was caused by a top-hat shaped jet,
and obtained the ratios, 
$R_1=E_p(\theta_v)/E_p(0)\sim10^{-2}$ and 
$R_2=E_{\rm iso,\gamma}(\theta_v)/E_{\rm iso,\gamma}(0)\sim10^{-4}$, 
for the narrow jet with fiducial parameters.
For off-axis jet emission, these values depend on the profile of angular distribution of the bulk Lorentz factor and 
 intrinsic emissivity near the periphery of the jet.
If the jet is structured like a Gaussian or power-law profile, then
$R_1$ and $R_2$ may be larger in the off-axis case \citep[e.g.,][]{Salafia2015}, so that on-axis $E_p(0)$ and $E_{\rm iso,\gamma}(0)$ may be smaller than the present estimates.
More quantitative arguments are beyond the scope of this paper.

\section{Discussion}

We have investigated an off-axis jet scenario in which we have invoked a two-component jet model to explain the observational results of GRB~190829A.
The best-fitted model in this paper is shown by solid lines in Fig.~\ref{lightcurve}.
According to our model, the early X-ray and optical afterglow was off-axis emission from the narrow jet, which may also be responsible for VHE gamma-rays detected at $\sim2\times10^4$~s, and the late X-ray and radio afterglow came from the wide jet (Figure \ref{fig:scenario}).
Since the narrow jet was viewed off-axis, the prompt emission was dim and soft due to the relativistic beaming effect.
On the other hand, the wide jet had the isotropic-equivalent kinetic energy $E_{\rm iso,K} \sim 10^{53}$~erg which was much larger than the observed isotropic equivalent gamma-ray energy $E_{\rm iso,\gamma} \sim 10^{49-50}$~erg.
If the wide jet has a typical value of the efficiency of the prompt emission, our result would become inconsistent with the observational result because it is seen on-axis. 
Since the initial bulk Lorentz factor of the wide jet is $\Gamma_0=20$, the jet is likely to be dirty (i.e., highly  loaded by baryons) and it may have a large optical depth.
It may be as small as $\eta_\gamma=E_{\rm iso,\gamma}/(E_{\rm iso,\gamma}+E_{\rm iso,K})\la10^{-3}$ unlike typical bright GRBs with high Lorentz factors.

We have also estimated the VHE gamma-ray flux at $2\times10^4$~s and have found that the narrow jet dominates  the observed gamma-ray emission. Since the synchrotron radiation is bright enough due to the large jet energy, the observed VHE gamma-ray flux, $\nu F_{\nu}\sim10^{-11}$~erg~s$^{-1}$cm$^{-2}$, is able to be expected by SSC mechanism.
In this paper, we independently calculate two emission components from two jets.
External inverse Compton with seed photons coming from the companion jet might  be effective \citep[e.g.,][]{Zhang2020b}.
Such an interaction between two jets remains to be future work.

There are still some observed components that are brighter than the prediction of our jet model.
They may be other components. For example, very early ($T\la7\times10^2$~s) X-ray emission should be the contribution from late prompt emission like flares.
Or if the jet is structured, the early X-ray afterglow shows a plateau phase or an additional peak
\citep{Beniamini2020a,Beniamini2020b,Oganesyan2020}.
The observed optical flux later than $\sim5\times10^4$~s is a supernova component \citep{Hu2020}.
At the late epoch ($T\sim10^{7}$~s), the 15.5~GHz radio flux also exceeds our numerical result, which could be other components such as counter-jet emission.

As seen in the right panel of Fig.~\ref{lightcurve}, our theoretical radio fluxes in both 1.3 and 15.5~GHz 
with fiducial parameters
sometimes overshot the observed ones. However, the excess is only within a factor of two, and this difference may come from the uncertainty of our simple model.
More realistic modeling may solve this problem. For example, structured jets such as Gaussian jets instead of uniform jets would decrease the radio fluxes keeping the X-ray and optical brightness unchanged \citep[e.g.,][]{Cunningham2020}.

Late-time ($T\sim10^{6-7}$~s) X-ray synchrotron emission from the wide jet
with fiducial parameters
is about a factor of two smaller than observed data (see the red solid line in the right panel of Fig.~\ref{lightcurve}).
In calculating the synchrotron radiation, we have assumed the Thomson limit to derive $\nu_c$ for simplicity. If we consider the Klein-Nishina effect \citep{nakar2009,murase2010,wang2010,murase2011,jacovich2020,Zhang2020b}, the Compton $Y$ parameter becomes smaller, so that $\nu_c$ becomes larger. 
Then, the X-ray flux increases if $\nu_c$ is around the X-ray band. As a limiting test case, we have calculated the X-ray synchrotron emission setting $Y=0$ all the time.
In this case, the X-ray flux actually becomes larger but by less than ten times. It is expected that inclusion of the Klein-Nishina effect causes the increase of the hard X-ray flux. Other possibilities to have a larger X-ray flux in the late epoch include delayed energy injection \citep[e.g.,][]{Zhang2006} and/or a low-energy part of SSC or external inverse-Compton emission \citep[e.g.,][]{Fan2008,Zhang2001ssc,Zhang2020b}.

The initial Lorentz factor of our narrow jet is $\Gamma_0=350$, which may be similar to or slightly smaller than those of long GRBs with VHE gamma-ray detection.
For GRBs~190114C and 180720B, the afterglow onset peak time may imply  the initial Lorentz factor of $\approx500$ and $\approx450$, respectively \citep{HuangXL2020}.
Furthermore, for both the narrow and wide jets, microphysics parameter $\epsilon_B$ is on the order of $10^{-5}$,
which is also similar to the other two long VHE events
\citep{Ajello2020,fraija2019a,fraija2019b,fraija2019c,wang2019,Jordana2020}\footnote{
GRBs with GeV gamma-ray emissions detected by {\it Fermi}/LAT may also have small values of
$\epsilon_B$ \citep{Beniamini2015,Tak2019} 
}.
At present, although the number of VHE events is small, these values are common for events with detectable VHE gamma-rays.
If there is no magnetic field amplification, $\epsilon_{\rm B}$ is about $10^{-7} (n_0/0.01{\rm cm}^{-3})^{-1} (B_{\rm ISM}/3{\rm \mu G})^2$, where $B_{\rm ISM}$ is the magnetic field strength for the ambient medium.
Therefore, the magnetic field in the emission region of those GRB afterglows is expected to be amplified. 
Although the mechanism has not been understood yet \citep[e.g.,][]{Tomita2019}, more detailed observations of VHE gamma-rays would provide us a new hint of the magnetic field amplification mechanism \citep[e.g.,][]{lemoine2015}.

The initial opening half-angle of the narrow jet is $\theta_0=0.015$~rad
as a fiducial value.
This is near the lower limit of previously measured values for long GRBs, however, it is still larger than the smallest one \citep{zhao2020a}.
In our model, the narrow jet is seen off-axis, resulting in dim prompt emission. Nevertheless, this event was observed since it occurred nearby.
Hence, similar but distant ($z\gg0.1$) events must be viewed on-axis to be detected.
However, a small solid angle of the narrow jet decreases the detection rate, which may explain the small number of VHE gamma-ray events that have been detected so far.

Compared with other long GRBs with radio detection, GRB~190829A showed a lower radio afterglow luminosity \citep{Rhodes2020}, 
which allows us to adopt a low ambient density, $n_0=0.01~{\rm cm^{-3}}$
as a fiducial value.
However, there may be two classes in long GRBs, radio-loud and radio-quiet events \citep{Zhang2020}.
Although radio-loud GRBs have slightly larger isotropic-equivalent energies
$E_{\rm iso,\gamma}$ of the prompt gamma-ray emission, the $E_{\rm iso,\gamma}$ distributions for the two classes look similar \citep[see Fig.~11 of][]{Zhang2020}.
It might be possible that long GRBs arise in the rarefied medium. Such an environment appears when the wind of a progenitor star is strong, or the bursts occur in the superbubble made by OB association.

\section*{Acknowledgments}
We thank Katsuaki~Asano, Kunihito~Ioka, Kazumi~Kashiyama, Takanori~Sakamoto, Motoko~Serino, Shuta~J.~Tanaka, and Kenji~Toma for valuable comments.
We also thank the referee
for his or her helpful comments to substantialy improve the paper.
This research was partially supported by JSPS KAKENHI
Grant Nos. 18H01232 (RY), 20H01901 (KM), 20H05852 (KM) and JP19H01893 (YO).
R.Y. deeply appreciates Aoyama Gakuin University Research Institute for helping our research by the fund.
The work of K.M. is supported by NSF Grant No.~AST-1908689.
Y.O. is supported by Leading Initiative for Excellent Young Researchers, MEXT, Japan.
%

\section*{DATA AVAILABILITY}

X-ray observations from {\it The Neil Gehrels Swift Observatory} are available at 
\verb|https://www.swift.ac.uk/xrt_curves/00922968/|.
Optical and radio data are taken from \citet{Chand2020} 
and  \citet{Rhodes2020}, respectively.
The theoretical model data underlying this article will be shared on reasonable request to the corresponding author.


\label{lastpage}

\onecolumn

\twocolumn

\end{document}